\documentclass[a4paper]{spie}  

\usepackage{amsmath,amsfonts,amssymb}
\usepackage{graphicx}
\usepackage[colorlinks=true, allcolors=blue]{hyperref}
\usepackage{multicol}
\usepackage{adjustbox}
\usepackage{hyperref}

\title{Neural networks applied to forecasting the astroclimatic and meteorological parameters in an operational context}

\author[a,b]{Camilo Weinberger}
\author[a,b]{Elena Masciadri}
\author[a,b]{Marlene de Sepibus}
\author[c]{Angel Otarola}
\author[c]{Stefan Mieske}
\author[d]{Christian Veillet}
\author[d]{Sam Ragland}

\affil[a]{INAF-OAA, Largo Enrico Fermi 5, 50125 Firenze}
\affil[b]{ADONI: Adaptive Optics National Laboratory in Italy}
\affil[c]{ESO, Alonso de Córdova 3107, Santiago, de Chile}
\affil[d]{LBTO 933 N. Cherry Drive, Tucson, AZ, USA, 85721}

\authorinfo{C.W.: E-mail: camilo.weinbergercerro@inaf.it\\  E.M.: E-mail: elena.masciadri@inaf,it}

\begin{document} 
\maketitle

\begin{abstract}
Forecasting optical turbulence and atmospheric parameters is crucial for science operations of present and new generation ground-based telescopes (ELT) as well as for free-space optical communication (FSOC). These forecasts are also useful to optimize the AO use, particularly in the ELT framework too. Recent research has demonstrated that the AR technique currently provides the best performances in predicting atmospheric and astroclimatic parameters on a time scale of 1 hour (e.g., RMSE 0.1" for the seeing), however deep learning offers some intriguing features that include the high adaptability, the correlation that can be found between different parameters treated simultaneously and different architectures can explore the capability for possible improvements.  For this reason, it is  important to investigate on the possibility to improve forecast performances beyond the score achieved by AR or simply to apply deep learning in those context in which it is not possible to apply the AR. In this contribution we investigate a set of algorithms belonging to the family of the neural networks that in principle are more suitable than the Random Forest (RF) model for this kind of application. We address specific questions related to the forecast performances in operational configurations. Besides that, we also investigate the impacts on gaps on the trained data set due to the lack of measurements of the optical turbulence during the day time. We show evidences that with measurements covering the whole 24-hours forecast performances might improve. Our test has been done on the wind speed in this contribution. 
This work is done in the framework of the operational forecast system for LBT (ALTA Center) and for the VLT (FATE project) that are leaded by our team.
\end{abstract}


\section{INTRODUCTION}
The forecast of the optical turbulence is a challenging goal for applications to the ground-based astronomy. For many years most of the scientific efforts of the scientific community have been concentrated on the demonstration of the possibility of forecast the optical turbulence with the mesoscale atmospheric model and with the parameterization of the optical turbulence in these models \cite{masciadri1999,cherubini2008,basu2020}. An important step ahead has been achieved later tackling the problem of the calibration of the mesoscale atmospheric models that greatly improved the model performances \cite{masciadri_jabouille2001,cherubini2008b,hagelin2011}. In the last decade the scientific attention is more focused on the demonstration of the possibility to provide operational forecasts with the final goal to support the science operations of the ground-based telescopes \cite{masciadri2013,masciadri2017,lyman2020}.  Our team is leading at present two operational forecast systems, more precisely the ALTA Center\footnote{\href{http://alta.arcetri.inaf.it}{http://alta.arcetri.inaf.it} - ALTA Center is a long term project.} above the Large Binocular Telescope (LBT) at Mt. Graham (Arizona, US) and the FATE\cite{masciadri2023} above the Very Large Telescope (VLT) at Cerro Paranal (Chile) that aim to provide optical turbulence (OT) forecasts, a critical factor for optimizing science operations of the two facilities. 
Recently, our team proposed an autoregressive (AR) method \cite{masciadri2020,masciadri2023} that is able to achieve an accuracy of  RMSE = 0.1’’ outperforming some machine learning approaches, including the Random Forest algorithm \cite{milli2019,turchi2022}. In the last years machine learning has been used in the field of the forecast of the OT even if with pretty different goals with respect to what we are analyzing here \cite{bolbasova2021,cherubini2022}. The interest for this topic is however growing on \cite{dasgupta2026,medlej2026}. Starting on those results, we have now extended our investigation to further algorithms to explore the capabilities of neural network (NN) models, expressly dedicated for time series, incorporating various filtering and optimization techniques to investigate the space of improvement upon previous benchmarks \cite{weinberger2026}. Neural networks and machine learning present indeed interesting features and great potentialities to improve forecast performances. A comparative analysis is performed across various models, including Random Forest, LSTM, TCN, and a sequence-to-sequence LSTM autoencoder. 
Moreover, in this contribution, beside to the investigation of the seeing forecast performances we try to answer to specific questions addressing the operational context: {\bf (1)} which is the optimized extension of the dataset over which to calculate the training to saturate the performances ?;  {\bf (2)} which is the dependency of results on training data-set period ? How do we need to upgrade a training data-set over the years ?; {\bf (3)} which is the possibility to predict the trend of the optical turbulence on a slightly longer time scales than 1h and shorter than 1 day (e.g 3 hours)?
All these questions address specific operative procedures.

\section{Methodology}
\label{method}

\begin{figure}
    \centering
    \includegraphics[width=0.65\linewidth]{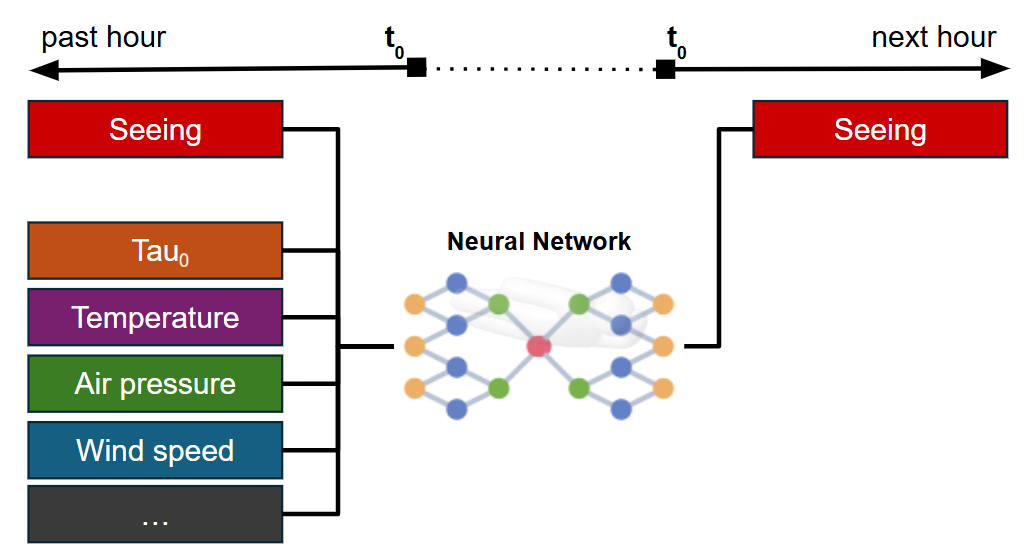}
    \caption{Schematic representation of the logical procedure followed by neural networks to provide an estimate of a parameter related to the future on the base of inputs belonging to the past.
}
    \label{fig:scheme}
\end{figure}
Figure \ref{fig:scheme} reports the schematic representation of the logic followed by neural network to provide the optical turbulence (OT) forecast. In this study, we evaluate the performance of some machine learning (ML) and Neural networks (NNs) models in forecasting the seeing above Paranal and Mt. Graham starting from the historical meteorological and astroclimatic parameters obtained by dedicated sensors in each location, more precisely the telemetry of the LBT for Mt.Graham and the ESO Ambient Conditions public repository for the VLT. These models contain multi parameters inputs and a single output (seeing at zenith). 
%
The selection of the parameters has been firstly done for the Paranal case. Due to the fact that the sensors and instruments for OT estimates above the two sites are different we tried to select the same parameters above the two sites and, where it was not possible, we tried to have at least a number of parameters as similar as possible in the training sample.
For Paranal we considered as input for the training: the seeing at zenith ($\varepsilon$),  the time coherence $\tau_0$, ground layer fraction (GLF), the air pressure, the temperature at 30m over the ground, the wind speed (3 components) and the target RA/DEC of the observed star. 
For Mt. Graham we considered as inputs for the training: the seeing corrected by the air-mass, the seeing,  the air pressure, the air temperature, the relative humidity, wind speed/direction (REAR), the wind speed/direction (FRONT). The terms `REAR' indicates the sensors located on the roof of the dome (back-position). The term `FRONT' indicates the sensors located on the roof of the dome (front position).
Additionally, for both sites, we included the cyclical encoding of time (cos(day/365), sin(day/365), cos(hour/24), sin(hour/24), the asymmetric moving average of the seeing in the last 10 minutes ($\varepsilon_{mean}$) and the asymmetric moving average of the standard deviation of the seeing in the last 10 minutes ($\varepsilon_{std}$).

To evaluate algorithms performance, the full 8 years period [2016-2024] with exception of year 2023 has been selected as a sample used for the training; data from the full year 2023 have been used for the testing. Data-set have been treated with a 2-minute resampling interval. To address the aforementioned research questions described in the Introduction (questions (1) and (2)), we investigated two different typology of forecasts:

\begin{itemize}
    \item \textbf{Case A}: the forecast of the average of the seeing within the next hour. 
    \item \textbf{Case B}: the forecast of the temporal evolution in the forthcoming next hour. This case treats forecast time series of one hour within 30 points (1 point every 2 minutes). Case B represents a bigger challenge in comparison of case A because we must predict the shape of the temporal evolution of seeing instead than the average, and in consequence the error increases. In other words, there is a  large number of different functions that could have the same average but different paths in the same interval. Both cases (A and B) are useful for flexible scheduling and science operations at short time scales.  
\end{itemize}
    
The models/algorithms used for this study are:
\begin{itemize}
    \item Random Forest (RF): it is a tree decision algorithm, this model is originally a classification mode, but as random comparator, it can be extended to forecast some points, in this case it compares each point of the time series to the forecast of the next step.
    \item Temporal Convolutional neural network (TCN): Convolutional neural network family is expressly dedicated to subtract specific  features from the time--series noisy segments to forecast the future time--series.
    \item Long-Short Term Memory (LSTM): this is a neural network dedicated to analyse complex time--series. Differently from TCN it takes into account long temporal series of the past to predict the next step.
    \item Long-short Term Memory sequence--to--sequence (LSTM seq-2-seq): this is a double version of LSTM. There are two models, where the first one encodes the data and the second one generates the time--series and optimizes the step--by--step inside the model.
\end{itemize}

Beside, in order to tackle question (3), we also investigated which is the forecast performance of the trend of the parameter (at present we are considering the seeing) on a few coming hours. To do this, we forecast the slope of linear regression of the time series for the next 1, 2 and 3 hours. As for the forecast of the average (case A) we provide the forecast of one value (i.e. a scalar) that tells us how much increases/decreases the seeing in a large time period in the future.

In this contribution we go to present a selection of results as the complete analysis will be shown in the paper Weinberger et al. \cite{weinberger2026}.  In other words we focus our attention on results reported in the poster and we refer the reader to a more completed paper to be published on a peer-reviewed journal. 

\section{Analysis and results}

\subsection{Question 1: optimal performances and optimal number of years necessary for the training}

To answer to question (1) we considered a set of test cases in which the training data-set is obtained by a number of years increasing from 1 to 7. Once the training done, we looked for the saturation level of forecast performances obtained by the testing sample. The sample of test is made by the whole year 2023. We considered the case A and B for the four algorithms described in Section \ref{method}.

\subsubsection{Case A}

\begin{figure}[h]
    \centering
    \includegraphics[width=0.65\linewidth]{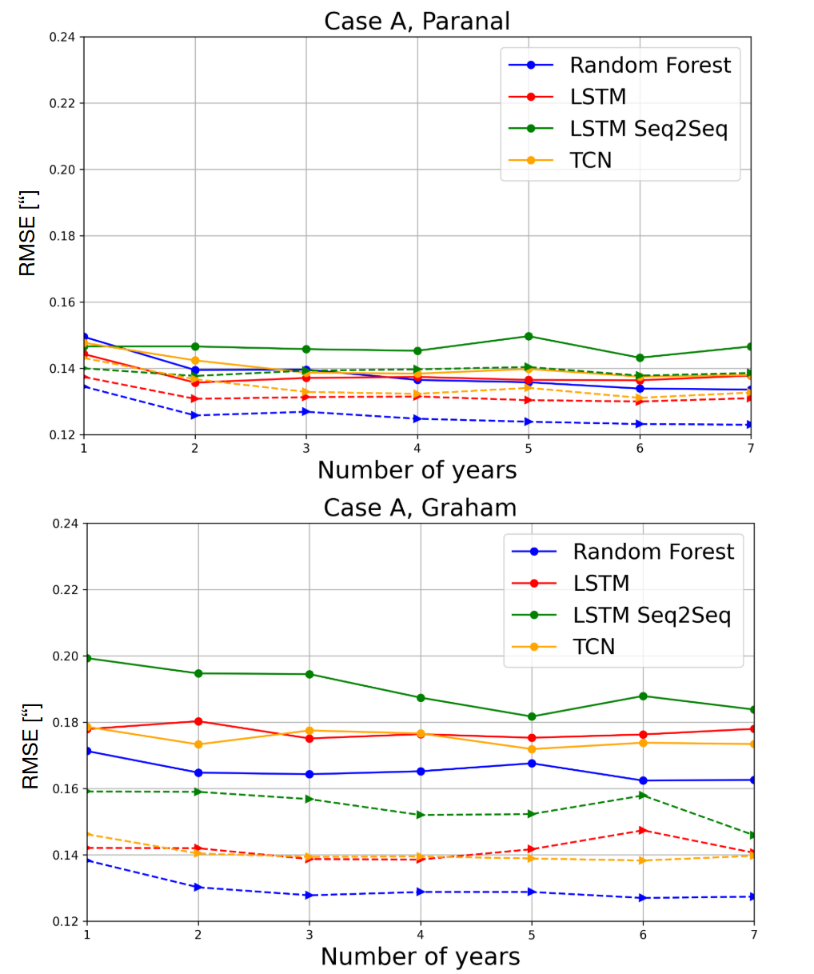}
    \caption{RMSE versus the number of the years considered in the sample of training data-set. The different colors indicate the different algorithms. The continuous lines represent the case of the total seeing while the dashed line the case in which the seeing $\varepsilon$ $<$  1.5 arcsec. }
    \label{fig:plot}
\end{figure}

 Figure \ref{fig:plot} shows the RMSE versus the number of years considered in the training obtained for the all the four algorithms above the two sites of Cerro Paranal (VLT) (Figure \ref{fig:plot}-top) and Mt.Graham (LBT) (Figure \ref{fig:plot}-bottom). Full lines indicate the case of total seeing, dashed lines consider the case of seeing $\varepsilon$ $<$ 1.5 arcsec. From these figures we conclude that the best algorithm seems to be the Random Forest (RF) for both sites. In the case of total seeing we achieve RMSE = 0.14 arcsec for Paranal and RMSE = 0.16 arcsec for Mt.Graham while in the case of seeing $\varepsilon$ $<$ 1.5 arcsec we achieve roughly  RMSE = 0.13 arcsec  above both sites.
 
 From Fig. \ref{fig:plot} we retrieve that, above both sites (Figure \ref{fig:plot}-top and -bottom), after an evident improvement obtained by passing from 1 year to 2 years, forecast performance achieve a saturation after roughly 2 years therefore very soon. This is true for basically all the algorithms. 
 For all the algorithms we observe, after a couple of years, a weak fluctuation of performances (a standard deviation $\sigma$ of around maximum 0.005 arsec as it shown in the Table \ref{tab:mean}.) 
 It is however possible to identify a few differences between the two sites of Paranal and Mt.Graham: 
 {\bf (1)} above Mt.Graham there is a much larger gain passing from the total seeing case to the seeing $\varepsilon$ $<$ 1.5 arcsec than above Paranal with similar performances in the case of seeing $\varepsilon$ $<$ 1.5 arcsec. 
 {\bf (2)} The best performances obtained in the case of seeing $\varepsilon$ $<$ 1.5 arcsec is very similar between the two sites of Mt.Graham and Paranal for the RF algorithm while for the other algorithms more important differences can be appreciated with better results above Paranal than above Mt.Graham. The nature of algorithms seems therefore to impact on the forecast performances. 
 {\bf (3)} for the same algorithm, the different performances are more evident in the case of the total seeing than in the seeing $<$ 1.5 arcsec.

\begin{table}[h]
    \centering
    \begin{tabular}{|c|c|c|}
    \hline
         & Paranal & Mt. Graham \\
         \hline
         Random Forest & 0.005 & 0.003\\
         LSTM & 0.003 & 0.002\\
         LSTM\_seq2seq & 0.002 & 0.006\\
         TCN & 0.004 & 0.002\\
         \hline
    \end{tabular}
    \caption{Fluctuation of the RMSE all along the total number of years considered for the training. Only the case of total seeing is shown. 
}
    \label{tab:mean}
\end{table}
.

\subsubsection{Case B}

We repeated here the estimations done for the case A using the same procedure but in the case of the forecast of the temporal evolution of the seeing. We refer the reader to the more completed version of the paper Weinberger et al.\cite{weinberger2026} for a complete analysis. Here we report in Table \ref{tab:raw} the values of the RMSE in the stable regimes for all the algorithms. We observe that, differently from the case A, here the forecast performances of NN are comparable to those of the Random Forest, and they are even better than RF for the case of $\varepsilon<1.5$ in Paranal, where TCN obtained the best performance of RMSE = 0.17 arcsec. This deserves a technical comment on the NN. We have seen indeed that in case A the RF is never overcome by the NN (in any conditions and in any site) while this is observed in case B, at least for the Paranal site. This seems to confirm that NN are more efficient than RF in those cases for which they are conceived to provide the best performances therefore for the time-series. Said that, probably the stability of the optical turbulence also plays a role as in the Mt.Graham case the RF stills show better performances. 

\begin{table}[h]
    \centering
    \begin{tabular}{|c|c|c|c|c|}
    \hline
         & Paranal &  & Mt. Graham & \\
         \hline
         & Full seeing & $\varepsilon<1.5''$ & Full seeing & $\varepsilon<1.5''$\\
         \hline
        Random Forest & 0.223 & 0.191 & \textbf{0.270} & \textbf{0.180}\\
        LSTM & 0.199 & 0.175 & 0.324 & 0.212\\
        LSTM\_seq2seq & 0.210 & 0.186 & 0.338 & 0.223\\
        TCN & {\bf 0.196} & \textbf{0.173} & 0.323 & 0.203\\
        \hline
    \end{tabular}
    \caption{Case B: values of the RMSE obtained in the conditions of stability for different NN algorithms, in the case of total seeing and seeing $\varepsilon$ $<$ 1.5 arcsec above Paranal and Mt.Graham. With `conditions of stability' we mean the time in which the saturation is reached. Values in bold style reports the best result for each category.}
    \label{tab:raw}
\end{table}

\subsection{Question 2: dependency of the forecast performances on the training period location}

We refer the reader to Weinberger et al. \cite{weinberger2026} for an exhaustive description of this topic. We simply say that we trained couple of consecutive years using always the same testing and we observe that we obtain the same level of saturation for the RMSE with a standard deviation of 0.001-0.002 arcsec passing from a couple to the other. This tells us that the dependency from the training data-set is weak and changing the couple of years does not lead to significant added values.

\subsection{Question 3: forecast of the trend/slope}

\begin{figure}[h]
    \centering
    \includegraphics[width=0.75\linewidth]{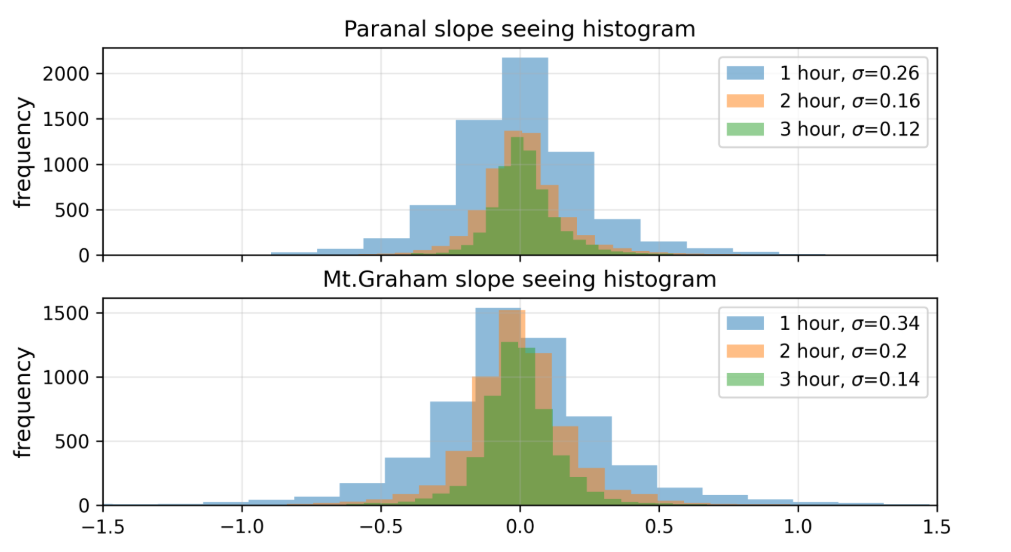}
    \caption{Histogram of seeing rate of change [''/h] for Paranal and Mt. Graham calculated on the sample of one solar year: 2023.}
    \label{fig:histogram}
\end{figure}

In this section we investigate the ability of these algorithms in forecasting the slope on a forecast time scale that is some how longer period than 1h and shorter than one day. We proceed therefore by studying the forecast of slopes for the case of 1h, 2h up to 3h to appreciate the behavior of the algorithms. We are considering the case A. We therefore predict the slop in the future by knowing the state of the slopes in the past. Figure \ref{fig:histogram} shows the histogram of the rate of change of the seeing expressed in (arcsec/h) calculated on a sample related to the whole 2023 year. On top the Paranal case, on bottom the Mt.Graham case. The different colors report the values for the cases of 1h (blue), 2h (orange) and 3h (green). The legend reports the value of the standard deviation of the histograms that, as expected, tends to decrease with the number of hours. In other words, the longer is the interval of time on which we calculate the forecast of the slope, the smaller is the standard deviation.  Table \ref{tab:trend}, shows the result of the standard deviation of the forecasts obtained with the three neural network algorithms (LSTM, LSTM seq2seq and TCN) obtained for a forecast time scale of one, two and three hours. 
Comparing Table \ref{tab:trend} with Fig. \ref{fig:histogram} we observe that the value of the standard deviation of the forecast of the slope reconstructs pretty well the value of the $\sigma$ retrieved from the histograms of observations for the respective time scales of 1h, 2h and 3h above the two sites even if the correlation is slightly better for the case of Paranal than for the Mt.Graham. We observe that the correlation scales with the time and we can even state that the best performances are obtained for the case of FTS = 3h. Said that at this time scale we have several values of the slopes that are infinitely small, almost equal to zero therefore predicting with good accuracy a value that is almost zero is not really useful. In conclusion, the resulting slope related to a long interval of time necessary approximates zero for raw estimates and for this reason it is not of great interest. 

\begin{table}[h]
    \centering
    \begin{tabular}{|c|c|c|c|c|c|c|}
    \hline
         & Paranal &  &  & Mt. Graham &  & \\
         \hline
         & 1H & 2H & 3H & 1H & 2H & 3H\\
         \hline
         LSTM & 0.27 & 0.17 & 0.12 & 0.39 & 0.22 & 0.16\\
         LSTM\_seq2seq & 0.29 & 0.16 & 0.12 & 0.39 & 0.21 & 0.16\\
         TCN & 0.27 & 0.16 & 0.11 & 0.38 & 0.18 & 0.13\\
         \hline
    \end{tabular}
    \caption{The standard deviation of the forecast of the slope (calculated on the sample of the 2023) as a function of the forecast time scale for the three different algorithms of neural networks.}
    \label{tab:trend}
\end{table}

\subsection{24H vs.night time training effect}

In this section we investigate what might be the effect produced by the gaps of measurements in the data-set sequence. We know indeed that, at present, we have not seeing monitors covering the day and night-time and the data-set covers only the night time. 
To test what might be the bias due to the lack of the information during the day we investigate the behavior of the RMSE (forecasts versus observations) obtained for an atmospheric parameter for which we can access night and day time observations, e.g. the wind speed (WS). We treat here the case (A) therefore the forecast of the average in the coming hour. Using the same testing sample selected during the night time, we compared the performances obtained with a training sample that takes into account night and day time observations (24H) and training data-set that takes into account the night time observation only (NIGHT). 

\begin{figure}[h]
    \centering
    \includegraphics[width=0.55\linewidth]{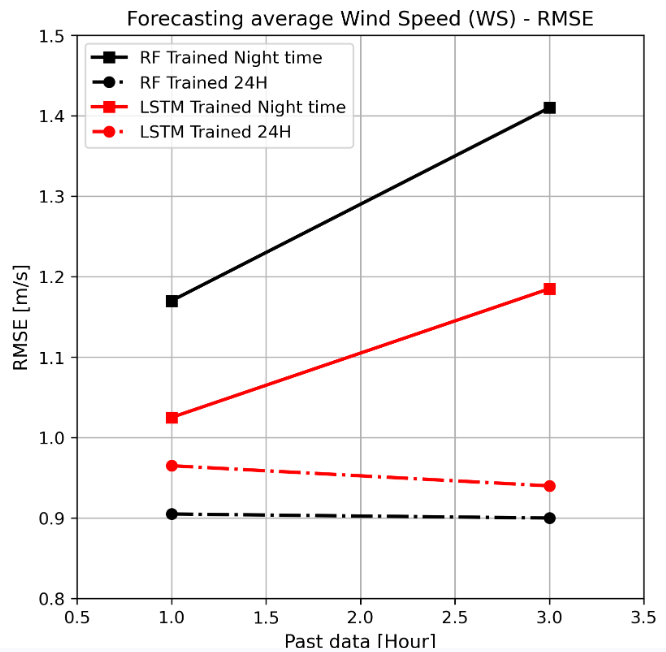}
    \caption{RMSE of the forecast of the wind speed (WS) versus the interval of time covering the past informations used by the training data-sets. The full and dashed line indicates the training data set done by night time only data-set and night and day time data-set respectively. }
    \label{fig:windspeed}
\end{figure}

Figure \ref{fig:windspeed} reports the RMSE obtained with the RF (black) and the LSTM (red) algorithms using a training calculated on 24H (dashed line) and on the night time only (full line). The sample of the training covers the period from 2016 to 2024 with the exception of the year 2023 that constitutes the testing sample. The RMSE is calculated using data-sets belonging to the past extended on a $\Delta$ = 1h and 3h. 
Looking at Fig.\ref{fig:windspeed} we conclude that: {\bf (1)} both algorithms obtained better performances if the training is done with observations belonging to the 24H than with the night only and this difference is  not negligible; {\bf (2)} if the training is done with data-set belonging to 24 hours, the longer is the $\Delta$(t) in the past, the better is the forecast accuracy for both algorithms; if the training is done with night time data-set only, forecast performances are worse for a longer $\Delta$(t) for both algorithms. This can be explained by the fact that in the latter case the number of data injected in the training is strongly reduced as the night is done of around 9 hours and the interval of time that we can cover with past data is much smaller in the case of 3 hours in the past with respect to 1 hour. Indeed, for the beginning of the night we have simply no data extended on the past 3 hours in the past;  {\bf (3)} if the training is done only with night time (NIGHT), the LSTM appears more performing than the RF while if the training is done with estimate of night and day time (24H) the RF is more performing than RF. That tells us that performances of the algorithm depend also on the parameter that we are treating. 

If we consider the case $\Delta$(t) = 3h the gain for RF in using the training on 24H instead of the night time only is of the order 55\%, for LSTM is of 30\% passing by a training done with night time data to a training done with night and day time. 

\begin{figure}[h]
    \centering
    \includegraphics[width=0.75\linewidth]{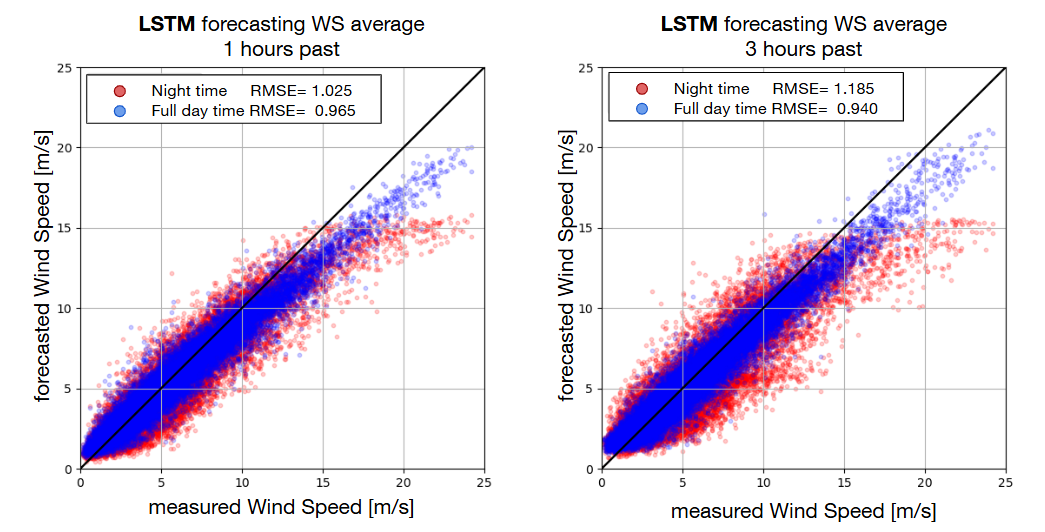}
    \caption{Scattering plot (and associated RMSE) between forecasted and observed wind speed using two different training data-set and the same testing data set.}
    \label{fig:scatter}
\end{figure}

In Fig.\ref{fig:scatter} is shown, for the LSTM case, the scattering plot of forecasted and observed wind speed obtained with the same testing sample but two different training data-set i.e. with night time dataset only (red) and with night plus day time data-set (blue). On the left is shown the case of a $\Delta$(t) = 1 hour in the past, on the right the case of $\Delta$(t) = 3 hours. We observe that the linear correlation decreases as the wind speed increases. In other words, using a training taking into account only night time data the algorithm fails to reconstruct wind speed higher than 15 m/s. Also we observe, as for Fig.\ref{fig:windspeed}, that the RMSE is definitely better when the training is done on the 24 hours. As a consequence the distribution of points around the bisector is narrower.  

We can not know if the optical turbulence behaves as the wind speed but there is a great probability that forecast performances improve if we use monitors covering the night and day time. For this reason our team decided to invest on a new instrument covering the 24H \cite{weinberger2026_2}.

\section{Conclusion}

In this contributions we treated different kind of neural networks together with the random forecast (RF) having a multiple goal: from one side we intend to quantifying the performances in predicting the seeing at short time scale (more precisely a forecast time scale of 1 hour) and on the other side we intend to answer to a set of questions related to the procedures to be used in an operational context. We analyzed the case of the forecast of the average of the seeing in the forthcoming hour (CASE A) and the forecast of the temporal evolution of the seeing in the forthcoming hour (CASE B) above the sites of the VLT (Cerro Paranal) and the LBT (Mt.Graham). 
Here the most relevant conclusions we achieved. A more detailed analysis will be reported in Weinberger et al. \cite{weinberger2026_2}:
\begin{itemize}

\item We proved that the optimal dataset length for the training to achieve stable forecast performances is equal to around two years. After that length the performances reach a sort of saturation. The fluctuation on the RMSE obtained considering longer period for the training is very small i.e. order of 0.002 arcsec.

\item We find that performances are weakly dependent on the algorithms even if there are some differences depending on the typology of the forecast. In case A the best performances are in any case achieved with the Random Forest (RF) for both sites. RMSE = 0.126 arcsec at VLT and RMSE = 0.130 arcsec at LBT, both values for the seeing $\varepsilon$ $<$ 1.5 arcsec. In case B the best algorithm is the TCN for the VLT while for the LBT the best algorithm is the RF. RMSE = 0.173  arcsec at VLT and RMSE = 0.180 arcsec at LBT, both for the seeing $\varepsilon$ $<$ 1.5 arcsec. Considering the forecast performances obtained by the reference models Precasting (fore the case A) and Prediction by Persistence (for the case B) (see de Sepibus et al.\cite{desepibus2026}) we can conclude that in the case A above both sites NN can achieve better performances than the reference even if the gain is small for VLT and more consistent for LBT. In case B, NN do not achieve performances better than the reference at VLT. The reference is overcome at LBT even if the gain is small. 

\item Using Random Forest we outperform results of  Milli et al., 2020; Turchi et al., 2022 if we consider the forecast of the average (case A). 

\item The forecast of the trend on the scale of 1h to 3h is well reconstructed by the algorithms tested here but the information is poorly useful as the slope on a those time scale is pretty close to zero. 

\item We showed that the training with 24H dataset instead of night only improves significantly the forecast performances for the WS. That might be the same for the seeing. We have a project/design for developing monitor covering the 24H and a request for funds has been done.

\end{itemize}

\acknowledgments 
 
We acknowledge the The European Southern Observatory (ESO) and the Large Binocular Telescope Observatory (LBTO) to support this study. This study has been carried out in the context of the the contracts  PO102958/ESO/20/95952/FLAB (ESO-INAF) and the contract ALTA 2: ENV002 (LBTO-INAF). 

\bibliographystyle{spiebib} 

\end{document}